\documentclass[reprint, onecolumn, amsmath,amssymb,jap,aip]{revtex4-2} 

\usepackage{graphicx}
\usepackage{dcolumn}
\usepackage{algorithm}
\usepackage{algorithmicx}
\usepackage{algpseudocode}
\usepackage{bm}
\usepackage{epsfig}
\usepackage{mathptmx}
\usepackage{times}
\usepackage{units}
\usepackage{amsmath}
\usepackage{upgreek}
\usepackage{tabularx}
\usepackage{todonotes}
\usepackage{threeparttable}
\usepackage[space]{grffile}
\usepackage[normalem]{ulem}
\usepackage{subfig}
\usepackage{multirow}
\usepackage{array}
\usepackage{xcolor}
\usepackage{soul}

\usepackage{caption}
\usepackage{physics}
\usepackage[T1]{fontenc}
\usepackage{courier} 

\begin{document}
\title{Fully Parallel Multi-Agent Photonic Optimizer}
\author{Ghazi Sarwat Syed}\affiliation{IBM Research -- Europe, S\"{a}umerstrasse 4, 8803 R\"{u}schlikon, Switzerland}
\author{Philipp Schmidt}\affiliation{Kirchhoff-Institut f\"{u}r Physik, Im Neuenheimer Feld 227, 69120 Heidelberg, Germany}
\author{Frank Br\"{u}ckerhoff-Pl\"{u}ckelmann}\affiliation{Kirchhoff-Institut f\"{u}r Physik, Im Neuenheimer Feld 227, 69120 Heidelberg, Germany}
\author{Jelle Dijkstra}\affiliation{Kirchhoff-Institut f\"{u}r Physik, Im Neuenheimer Feld 227, 69120 Heidelberg, Germany}
\author{Wolfram H.P Pernice}\affiliation{Kirchhoff-Institut f\"{u}r Physik, Im Neuenheimer Feld 227, 69120 Heidelberg, Germany}
\author{Abu Sebastian}\affiliation{IBM Research -- Europe, S\"{a}umerstrasse 4, 8803 R\"{u}schlikon, Switzerland}\affiliation{Kirchhoff-Institut f\"{u}r Physik, Im Neuenheimer Feld 227, 69120 Heidelberg, Germany}

\maketitle

\section*{Abstract}

\noindent{\textbf{Optimization problems are central to many important cross-disciplinary applications. In their conventional implementations, the sequential nature of operations imposes strict limitations on the computational efficiency. Here, we discuss how analog optical computing can overcome this fundamental bottleneck. We propose a photonic optimizer unit, together with supporting algorithms that uses in-memory computation within a nature-inspired, multi-agent cooperative framework. The system performs a sequence of reconfigurable parallel matrix–vector operations, enabled by the high bandwidth and multiplexing capabilities inherent to photonic circuits. This approach provides a pathway toward fast paced and high-quality solutions for difficult optimization and search problems.}}

\begin{flushleft}
 \textbf{Keywords}: Optimization Problems, Nature Inspired Computing, Integrated Photonics
\end{flushleft}

\section*{Introduction}

\begin{figure}[h!]
    \includegraphics[width=\textwidth]{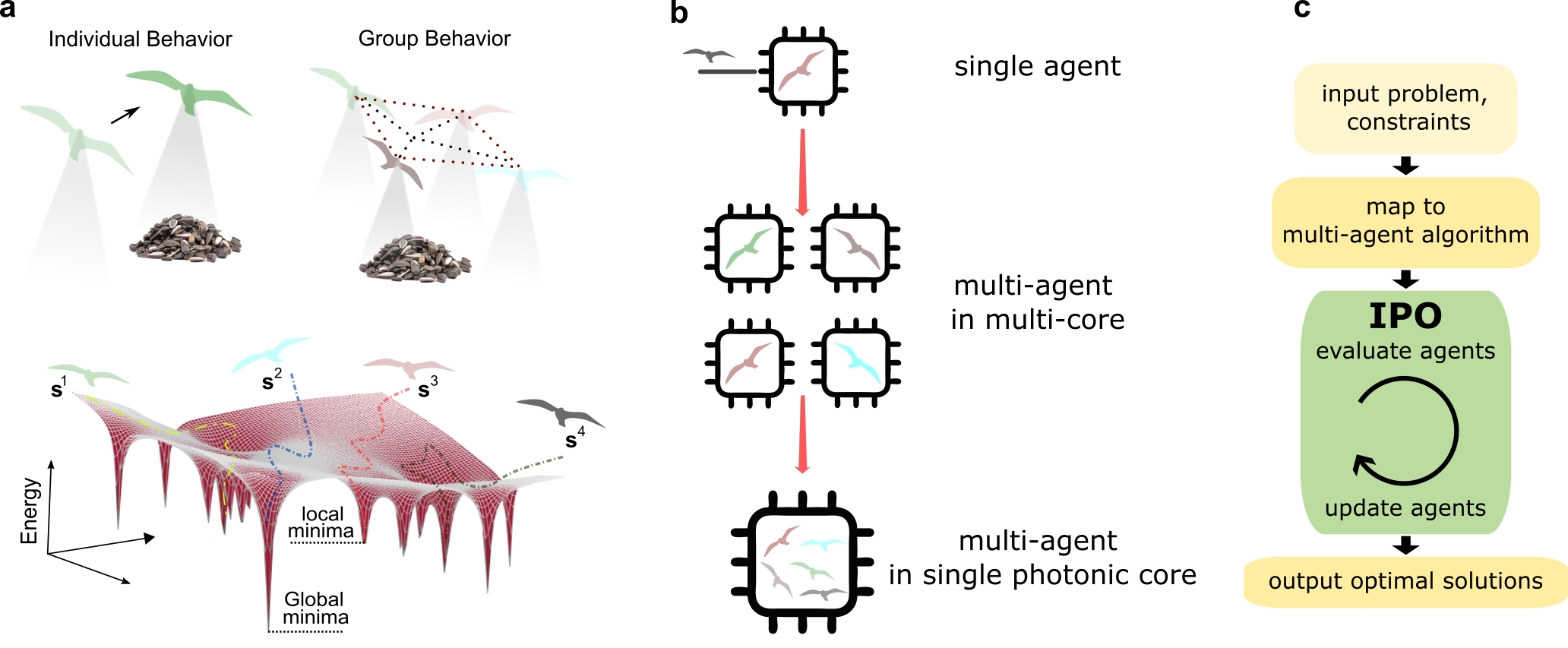}
    \caption{\textbf{Concept of Photonic Optimizer}. \textbf{(a)} Illustration of nature inspired multi-agent optimization. A foraging avian must explore ($\vec{d}
$) the landscape using their sensory range. The likelihood of finding the optimal food resource is improved through cooperation with a population of avians. When translated to the energy landscape of a typical NP-hard optimization problem, this strategy, leveraging the individual and social characteristics of multiple independent agents, can enable convergence to the global optimum in difficult optimization tasks. \textbf{(b)}  Comparison of digital electronic architectures with the proposed photonic optimizer for solving multi-agent optimization problems. Digital electronics require many sequential processing steps, which can be executed on a single core or distributed across multiple cores. In contrast, the entire problem can be processed within a single photonic optimizer in a truly parallel, in-memory computing manner. \textbf{(c)} Flowchart describing the operational flows in the proposed integrated optimizer unit. When properly defined through chosen constraints, loss functions and update rules, the optimizer can iterate over a problem with multiple agents, evaluating all agents in a single cycle.}
    \label{fig:1_concept}
\end{figure}

\noindent In the nature, many species engage in social associations for a variety of reasons. One of the most common being social foraging, which is the process by which individuals in a group work together to forage for resources (\onlinecite{kennedy1995particle,dorigo2006ant,coopertaiveschloesser2021individual,torney2011signallingcooperative}). Such cooperation confers significant mutual benefits, including improvements in search efficiency and the dissipation of risk factors. Key to the success of this tactic lies predominantly in an improved exploration and exploitation balance that is achieved through the social and cognitive decision-making abilities of the interacting individuals (see Figure \ref{fig:1_concept}a). Several algorithms in computational science take inspiration from such ``cooperative’’ behavior to solve difficult mathematical problems. Broadly, these include algorithms that implement cooperative artificial intelligence  (\onlinecite{dafoe2021cooperative,gupta2017cooperative,reinforcetan1993multi,booktalbi2009metaheuristics,surveygogna2013metaheuristics,blum2003metaheuristics}) and solve optimization problems. Multi-agent-reinforcement learning and metaheuristic algorithms such as ant colonies and particle swarms are some examples, which implement decentralized collaboration and coordination among synthetic individuals (agents) to accomplish their tasks. When applied to deep neural networks and \textit{np} hard optimization problems, such algorithms have resulted in impressive results, such as emulating the game of soccer (\onlinecite{dafoe2021cooperative} and solving combinatorial (\onlinecite{blum2003metaheuristics}) problems. \\

Nonetheless, such results continue to be based on the digital von Neumann computing systems, where data needs repeated shuttling between the processor and the memory units, on power-hungry interconnects (\onlinecite{sebastian2020memoryAbu}). This presents a limit on the achievable compute latency and incurs notable energy costs. In addition to this inherently rooted complexity, such computing systems can allow only sequential implementation of cooperative algorithms. Crucially implying that each agent must be processed individually, and the combined data further processed. This results in the real-time interaction aspect and learning being lost. \textcolor{black}{We also note that, in contrast to deeply pipelined workloads such as AI inference, where throughput is maximized by batching independent inputs, optimization problems exhibit an inherently sequential data flow. As a result, throughput is not the dominant performance metric. Rather, the total time to solution is governed by the accumulated per-iteration delays, such that reducing the latency of each iteration directly translates to lower end-to-end latency.} Whilst various custom silicon computing hardware (i.e. ASICs and GPUs) have been developed to improve these aspects (\onlinecite{zhou2009GPU,chakroun2013combiningGPU,tsukamoto2017acceleratorFPGA,Y2021gpuzhang}), they still depend on the same underlying electronic components, which are fundamentally limited by the electronic physics of Joule heating, RF crosstalk, capacitance, and by sequential bus-based data addressing architecture, which constraints data movement efficiency from limited bandwidths and cache coherency. For these reasons, there are strong incentives for developing \textcolor{black}{processing-in-memory} accelerators (\onlinecite{mutlu2022modern,Y2019computationalsebastian}) and supporting algorithms that allow for rapid solving of difficult problems in a manner that is rigorous and nature-inspired on hardware (see Figure \ref{fig:1_concept}b). \\

Here, we propose just such a framework in the form of an integrated photonic optimizer (IPO), \textcolor{black}{which is an iterative solver}. We aim to solve continuous and discrete optimization problems using the proposed IPO. Central to our approach is the exploitation of physical properties of integrated photonic circuits and devices (\onlinecite{rios2015integrated,wuttig2017phase}), along with the wavelength division multiplexing (WDM) capabilities of optics (\onlinecite{Y2021feldmannparallelconvo,ghazi2022integrated}), to perform multiply-and-accumulate (MAC) operations efficiently. These computations are executed in-place and in parallel. \textcolor{black}{By formulating the optimization problem as iterative analog MAC operations, we facilitate parallel exploration of the objective (cost) function landscape, targeting the sub-nanosecond latency achievable in optics (\onlinecite{hua2025integrated}}). In this scheme, agents are uniquely encoded into non-overlapping wavelengths of light. Each wavelength independently explores the solution space, shaped by the programmable configuration of the compute elements. This process occurs simultaneously across all agents and is guided by cooperative dynamics inspired by nature (see Figure \ref{fig:1_concept}b). Unlike traditional electronic architectures, where each agent is computed in separate hardware, our photonic system favors solutions that minimize information gain, thereby improving the chances of finding the global optimum. Moreover, by performing computation directly where data is stored, this approach bypasses the memory wall bottleneck and overcomes the speed limitations of electronic components.

\section*{Results}

\subsection*{Formal Definitions:}

\noindent Computing optimizations are common computational operations in which the goal is to find the best solution from all feasible solutions with in the configurational search space. In formal terminology, an optimization problem can be defined in terms of a tuple $S,f,r)$. Here, $S$ represents the search (state) space of the problem, $r$ the set of the problem's constraints, and  $f$ is the objective function. The objective function guides the search process, such that $f:S\rightarrow \mathbb{R}^m$. In most problems, the directive is either to minimize or maximize the objective function, i.e. $\underset{\textbf{s} \in S \!}{\text{minimize}} f(\textbf{s})$ or $\underset{\textbf{s} \in S \!}{\text{maximize}} f(\textbf{s})$, and $\textbf{s}$ is a vector, whose elements represent a decision variable that can be continuous or discrete. When the variables are discrete, the optimizations become discrete optimization problems. These are computationally harder to solve because an optimal solution has to be identified from a finite set of solutions. Integer and combinatorial optimization problems are examples of such problems, which are relevant in several technologically and scientifically important applications. The choice of an objective function is crucial in determining the quality of a solution,  and while often given for integer type problems, must be carefully devised for the combinatorial problems. Canonical examples include Lyapunov functions, such as those associated with the Ising and Hopfield models. \\ 

When solving such problems, the naive approach is to randomly initialize a candidate query that performs search operations "locally" on the solution landscape. During this process, the solution evolves to provide an improvement to the solution until the optimal answer or a convergence criterion is met. In cases where the landscape profiles are complex, the solutions can easily get stuck in the local minima. A possible analogy can be drawn from the descriptions of foraging by an avian predator (see illustrated in Figure \ref{fig:1_concept}a). With a limited sensory range, through movements, a solitary bird scans a terrain while foraging. When the terrain is large and complex, chances are the bird may converge to no gains (improvements), or a sub-optimal food resource (local minima). Such predicaments are avoided when foraging in large flocks.  In a flock, birds can discreetly examine different regions of the terrain, and through sensory communication choose to relay the information to the entire swarm. In effect, this scheme provides a means to "globally" search the landscape, thus, improving and speeding up the chances to locate resources.\\

In Figure \ref{fig:1_concept}a, we illustrate a hypothetical solution landscape to an objective function. Multiple attractor points provide local (sub-optimal) solutions. At any time instance, multiple disparate queries $\{\textbf{s}^\text{1},\textbf{s}^\text{2},...,\textbf{s}^\text{n}\}$ in the state space are allowed to scan the landscape. Each query belongs to a discrete agent described by a tuple $(\textbf{s},\vec{d}, F)$, where $F$ is the feedback given to an agent by the ensemble and $\vec{d}$ is its action. The exact implementation of these quantities depends on the chosen algorithm and we will elaborate on these terms in the following sections. It is suffice to suggest that an agent interacts with the environment and other agents in steps in time ($t$), to determine $ \vec{d^{t}}= f(F)$. Such a scheme makes it possible that the agents, independently of their starting points, can converge to better solutions through dislodging from parasitic local minima. An IPO unit can realize this functionality (Figure \ref{fig:1_concept}c). Specifically, it enables a multi-agent solver in which multiple agents, mapped to independent input vectors, explore the same cost landscape through matrix–vector operations. A set of update rules, which are defined by curated  algorithms guide the agents toward convergence to near-optimal solutions (see Figure \ref{fig:2_prop_IPO}a). 

\subsection*{Conceptual Hardware:}

\begin{figure}
    \centering
    \includegraphics[width=\textwidth]{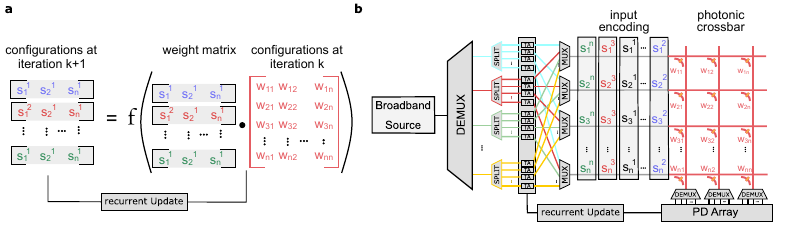}
    \caption{\textbf{Conceptual Integrated Photonic Optimizer Unit}. \textbf{(a)} In an IPO unit, every agent is encoded as an input vector carried on a unique wavelength (optical carrier). The optimization problem is then represented and solved through matrix–vector operations, with the matrix elements encoded in the states of photonic memory devices. Multiple agents can be processed in parallel, leveraging both in-memory and state update computations to efficiently explore the solution space. \textbf{(b)} Conceptual schematic of the compute core. The core consists of multiple individually addressable wavelength channels (one per agent) and a high optical bandwidth linear processor. After demultiplexing a broadband light source (e.g., ASE source, laser array, or frequency comb), each wavelength channel is split into $m$ distinct paths, each modulated by a tunable attenuator (TA) (e.g., EAM). This allows an input vector of length $m$ to be encoded on each channel. The channels are then multiplexed into the $m$ input waveguides of a photonic crossbar array, with each intersection also loaded with a tunable attenuator (e.g., phase-change material, EAM). The crossbar performs multiply-and-accumulate operations between matrix elements $w_{ij}$ and all frequency channels in a single pass. The resulting vectors of are measured via demultiplexed photodetectors. The outputs of the photonic agent evaluation are then processed according to the chosen algorithm to update the inputs for the next cycle.}
    \label{fig:2_prop_IPO}
\end{figure}

\noindent We instantiate an optimization problem to use iterative MAC operations, between a $m \times n$ matrix that maps the objective function and the problem constraints, and the input vectors $\textbf{s}^i$ that encode the state space. Figure \ref{fig:2_prop_IPO}b illustrates the proposed IPO unit.  Here, each agent is assigned a distinct wavelength channel. The unit has a compute core which takes advantage of data stationarity by keeping MAC 
computation and data in the same place. This is achieved by storing the numeric values of a matrix directly in tunable attenuators (reconfigurable and analog memory cell). Potential realizations for these attenuators are for example electric absorption modulators (EAM) or phase change material (PCM) patches  (\onlinecite{dong2024partial,Y2019fastXuanli,Y2021feldmannparallelconvo}). These memory cells are arranged at each intersection of a photonic crossbar circuit. Each memory cell performs a scalar multiplication operation of the form $\nu_{\text{out}}=w\times \nu_{\text{in}}$, where $w$ is the transmission state of the cell and $\nu_{\text{in}}$ is the input optical signal (\onlinecite{Y2021feldmannparallelconvo}). The summation of the individual products occurs in the photodetector at the end of each waveguide. The input vector, representing an agent (\(\vec{s}_\text{i}\)), is encoded in the amplitude of the optical signals sent to the input waveguides of the crossbar.  De-multiplexed input encoding and detection, together with the large optical bandwidth allow all agents to be processed in parallel. Moreover, a recurrence exists in which the agents, updated from the output of the feedforward MAC operations, are fed back as inputs for the next iteration.\\

Importantly, wavelength-division multiplexing (WDM) enables multiple input vectors or agents to be encoded in the intensities of distinct, non-overlapping wavelength channels. These channels can propagate through the compute core simultaneously, allowing several multiplication operations to be performed in parallel. A large number of agents can be generated from broadband light sources such as amplified spontaneous emission (ASE) sources, frequency combs  (\onlinecite{fortier_20_2019}), or laser arrays. Photonic crossbars generally require each input waveguide to be mutually incoherent with all others to prevent interference at the detector. Consequently, the proposed IPO would require two cascaded demultiplexers and a total of $n_\text{agents} \cdot n_\text{inputs}$ non-overlapping spectral lines when coherent sources are used. Alternatively, inherently incoherent light, such as that provided by ASE sources, can be employed. In this case, the second demultiplexer can be replaced by a symmetric splitter, as illustrated in Figure~\ref{fig:2_prop_IPO}b. After encoding, the channels are multiplexed into the crossbar and demultiplexed prior to readout by a photodetector (PD) array. While ASE light introduces significant inherent intensity noise, this can be mitigated to some extent by adjusting the optical bandwidth, optical power, or through averaging  (\onlinecite{bruckerhoff-pluckelmann_probabilistic_2024}). However, randomness is an intrinsic feature of some optimization algorithms, and in such cases, we can harness this noise to our advantage.\\

Mathematically, at each iteration $k$, the agents in the IPO collectively perform the operation $
\mathbf{s}_{k+1} = \mathbf{W} \, \mathbf{s}_k + \boldsymbol{\eta},$ where $\mathbf{s}_{k+1}$ denotes the recurrently updated state vector obtained from the previous state $\mathbf{s}_k$. Here, $\mathbf{W} = [w_{ij}]$ is the weight matrix, $\mathbf{s}_k = [\, s_k^{1}, s_k^{2}, \ldots, s_k^{J} \,]^T$ is the input state vector across agents at iteration $k$, and $\boldsymbol{\eta}$ represents additive noise. Crucially, all these operations are performed in parallel, eliminating also the need for explicit summation operations.

\subsection*{Experimental Validation:}

\begin{figure}
    \centering
    \includegraphics[width=\textwidth]{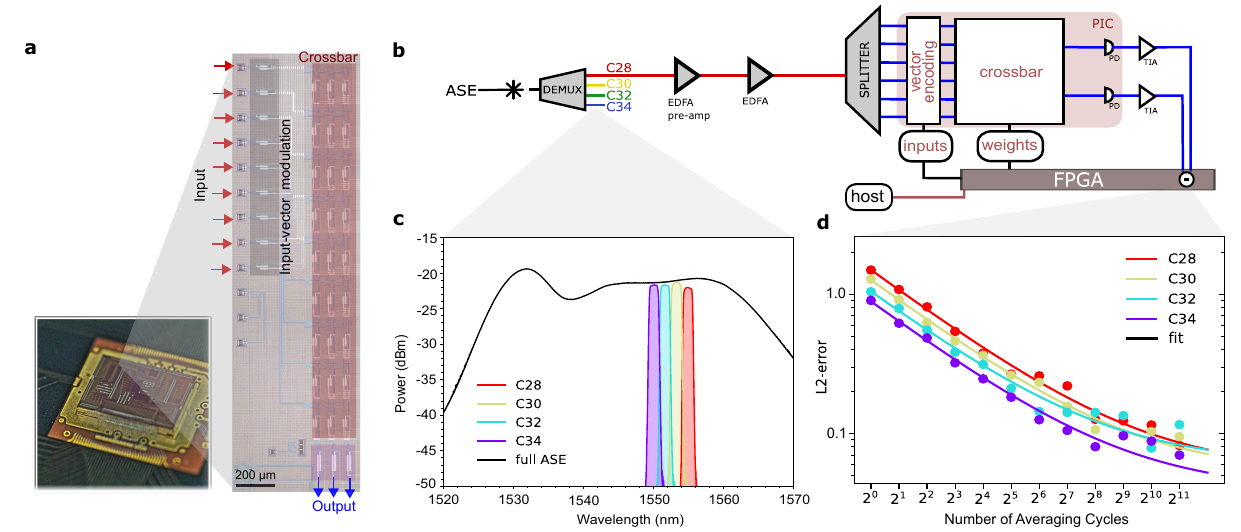} 
    \caption{\textbf{Prototype Integrated Photonic Optimizer Unit}. \textbf{a} Micrographs of a demonstrator silicon PIC. Various on-chip components are highlighted. \textbf{b} The IPO is evaluated for functionality and performance through a prototype hardware. Four distinct C-band wavelength channels (200GHz ITU grid channels C28, C30, C32 and C34) are spatially multiplexed to perform matrix-vector operations. A high-speed FPGA is interfaced with the photonic chip to handle auxiliary operations. Light from a broadband ASE source is split by a demultiplexer into the select channels, each subsequently amplified in two stages to roughly \unit[100]{mW} before entering the silicon PIC. The PIC incorporates programmable input weights along with a 9×3 programmable crossbar array. Photocurrents from the on-chip photodetectors are amplified by off-chip TIAs and fed back into the same FPGA, which also controls the input signals and matrix weighting. \textbf{c} A plot depicting the spectrum of the ASE source along with the four distinct 200 GHz channels. \textbf{d} A plot showing L2-error of matrix-vector multiply operations for each input channel across averaging, with fitted noise profiles obtained through average error of 1280 random input vectors length 64 and a constant weight matrix.}
    \label{fig:3_msm_setup}
\end{figure}

\noindent So far we have discussed a generic IPO that provides a compact means of processing multiple agents with real-time cooperation within a single processing core.  We now present a proof-of-concept experiment to evaluate the feasibility of the key functionalities required in the IPO. The setup (see Figure \ref{fig:3_msm_setup}a-c) consists of an ASE source emitting broadband chaotic light over the 1530–1560 nm range. This light is spectrally filtered by a wavelength-division multiplexer into four 200 GHz bandwidth channels, centered on specific ITU C-Band wavelengths: C28 (1554.94 nm), C30 (1553.33 nm), C32 (1551.72 nm), and C34 (1550.12 nm). The filtered light from a single channel is then amplified in two stages and injected into a integrated photonic circuit (PIC) consisting of a 9$\times$3 crossbar array containing electronic absorption modulators (EAM) at the cross point junctions. It also features EAMs on the input waveguides as well as integrated PDs at the output. The PD signal is amplified by off-chip TIAs and then fed to ADC ports of a FPGA board. The same board is also used to interface with the weight and input EAMs and thus constitutes the control unit of this experiment. \\

The optical noise arising from the ASE as well as the electrical noise injected by the receiver components lead to noisy MAC operations. We can reduce this accumulated noise via averaging multiple iterations of the same matrix-vector multiply (MVM) operation. We approximate the combined noise distribution as gaussian and thus are able to generalize the dependency of the noise on the amount of averages $n_{avg}$ with the form, $f(n_\text{avg}) = \frac{A}{\sqrt{n_\text{avg}}} + B$, where \( A \) and \( B \) are fitting parameters that determine the initial noise amplitude and residual ground error, respectively. We benchmark the system by performing matrix-vector multiplications with random input vectors and constant matrix weights.  The resulting output vector $\vec{y}$ is compared against the calculated ground truth using the $L_2$ error $\frac{\| \vec{y}_{\text{measured}} - \vec{y}_{\text{calculated}} \|_2}{\| \vec{y}_{\text{calculated}} \|_2}$. Figure~\ref{fig:3_msm_setup}d illustrates the averaged $L_2$ error values across the independent input channels, representing the injected noise that can be reconfigured through averaging.  In the following sections, we use the data obtained from these measurements to evaluate the feasibility of our proposed IPO unit. 

\subsection*{Algorithmic Strategies on the IPO Platform:}

\noindent In multi-agent optimization, each agent's behavior is influenced by a social interaction function \( F \). During optimization, the system converges to a state that minimizes the net loss. Feedback can be applied either deterministically or via stochastic and evolutionary changes. When no feedback is present between agents (\( F=0 \)), the process is deterministic and corresponds to \textit{sampling}.  The manner in which feedback is implemented enables novel algorithmic advances. We introduce a general \textit{dynamic referencing algorithm} (see Algorithm \ref{alg:dyn_ref}). The algorithm offers a candidate solution \( \textbf{s}^\text{c} \in S(t) \), defined as $\textbf{s}^\text{c} = \arg\min_{\textbf{s} \in S(t)} f(\textbf{s})$ at time step \( t \). The objective is to apply controlled random perturbations or operators to configurations other than \( \textbf{s}^\text{c} \), such that they differ minimally from \( \textbf{s}^\text{c} \) depending on the nature of $F$. These perturbations are designed to balance exploitation of the already obtained knowledge by the agent ensemble and exploration of individual agents independent of the combined knowledge. They can change over time adjusting the this balance. 
As a convergence criterion we introduce a similarity metric between \( \textbf{s}^\text{c} \) and the other configurations. For discrete inputs, this similarity can be, for example, the Hamming similarity \(\text{Sim}_{\text{Hamming}}(\mathbf{s}^{\text{c}}, \mathbf{s}^{\text{n}})\). This approach allows multiple input configurations to converge toward a single, high-confidence solution. There is a one-to-one correspondence between the number of agents (or wavelengths) and the solution quality: a greater number of agents enables broader exploration of the solution landscape, improving the chances of locating optimal regions and boosting convergence likelihood. Moreover, this aligns with the idea that confidence scales inversely with the Shannon entropy (\onlinecite{Y1948shannonmathematical}), defined as $E_\text{S} = -p \log p - (1 - p)\log(1 - p)$ where \( p = \frac{N_c}{N}, \quad 1 - p = 1 - \frac{N_c}{N}\). Here, \( N \) is the total number of parallel queries at a given iteration, and \( N_c \) is the count of configurations matching the candidate solution \( \mathbf{s}^{\text{c}} \). As more configurations align with \( \mathbf{s}^{\text{c}} \), the entropy \( E_\text{S} \) decreases, indicating growing consensus and confidence in the solution.\\

\begin{algorithm}[H]
\begin{algorithmic}[1]
\caption{Dynamic referencing algorithm}\label{alg:dyn_ref}
\State \textbf{(Best Selection)} Identify current best: \(\mathbf{s}^{\text{c}} \leftarrow \arg\min_i E_i\)
\State \textbf{(State Adaptation)} Apply every \(k\) iterations:
    \State \hspace{1em} For each agent \(\mathbf{s}^{\text{i}} \ne \mathbf{s}^{\text{c}}\):
    \State \hspace{2em} \textbf{Update:} apply deterministic Update
\State \textbf{(Optional Enhancements)} Apply as needed:
    \State \hspace{1em} \textbf{Crossover:} exchange information with another agent
    \State \hspace{1em} \textbf{Annealing:} accept probabilistic pertubation
    \State \hspace{1em} \textbf{Elitism:} exclude top agents from update
\State \textbf{(Energy Update)} Recompute \(E_i \leftarrow f(\mathbf{s}^{\text{i}}\)), refresh \(\mathbf{s}^{\text{c}}\)
\end{algorithmic}
\end{algorithm}

We now discuss some schemes to implement the perturbations. The first is the case of photonic, multi-agent optimization analogous to simulated annealing (\onlinecite{van1987simulatedannealing,booktalbi2009metaheuristics}. In this approach, injection of uncorrelated fluctuations on either the state itself or the deterministic update rule, enabled by a reconfigurable $\eta$ can dislodge agents into probable state space from parasitic minima.  A key requirement for this method is high-bandwidth random number generation and a decay profile applied to $\eta$ which, in our implementation, is naturally provided by the analog noise inherent to the IPO hardware. We combine this with the evolutionary principle~ (\onlinecite{muhlenbein1988evolution, katoch2021reviewevolution}) of crossover ($\bigoplus_f$) where in each cycle the agent inherits part of the best current solution. 
Our second approach implements particle swarm optimization (PSO), where agents explore the loss landscape based on physical principles of velocity and interaction, updating their trajectories based on their own and the ensemble experience  (\onlinecite{shi2001particleswarm,wang2018particleswarm}). In the following section, we emulate these algorithmic features under realistic conditions. Although the examples we present are not exhaustive, they are meant to illustrate and benchmark the unique capabilities enabled by the IPO unit.

\subsection*{Solving Discreet Optimization:}

\begin{figure}[h!]
    \includegraphics[width=\textwidth]{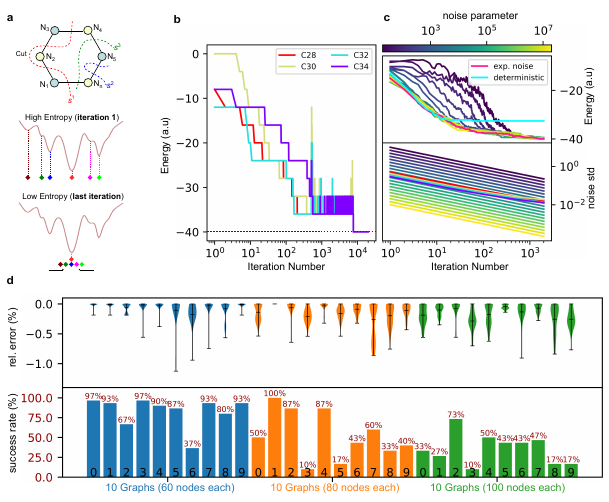}
    \caption{\textbf{Combinatorial Optimization Problems}. (a) The top panel presents a schematic of the Max-Cut graph optimization problem solved by multiple agents. The lower panels show the energy landscape under the dynamic referencing algorithm, where multiple agents (colored dots) evolve in parallel and converge toward the global minimum, progressively reducing entropy. (b) Exemplary emulated optimization of a circular graph of 40 nodes using measured compute accuracy levels for the four ITU-channels from the benchmark measurement. The dashed red line cuts the graph into two sets of complementary nodes. The edges are encoded as matrix weights in the photonic crossbar. \textbf{(c)} Top Panel: Average global energy of twenty runs discovering the Max-cut of a circular graph of size 40 for different decaying noise profiles. The cyan line shows the average for purely deterministic calculations. The red line the performance using experimentally measured noise levels. While deterministic multi-agent approach can enable optimal convergence with increasing number of agents and iterations, we find optimal noise profiles to allow quicker convergence to the optimal solutions. \textbf{(d)} Performance Benchmarks on non circular graphs. Examples are taken from the Biq-Mac library  (\onlinecite{biqmac}). \textcolor{black}{We use the experimentally measured noise profiles and four agents to solve each graph across 30 runs with different initial conditions. The success rate and relative errors are then computed from the fraction of runs that converge to the known solution. Even without deliberate hyperparameter tuning, every graph can be solved, with at least one successful trial in each case}. In the top panel, the bars indicate the maximum, minimum, and mean of the relative error.}
    \label{fig:5_max_cut}
\end{figure}

\noindent As a first example, we discuss the case of discrete optimization problems. We solve the constrained optimization graph problem of Max-cut (\onlinecite{Y2003MAXCUT,Y2011combinatorialMAXcut}. Max-cut (see Figure \ref{fig:5_max_cut}a) is typically utilized in integrated-circuit designing and imaging, and the goal is to partition a graph's vertices into two complementary sets (highlighted in different colors), such that the number of edges linking the two sets is as large as possible. Stated formally, given an undirected graph $G(v,e)$, where $v$ and $e$ are the graph's vertices (nodes) and the edges (interconnections) between nodes, respectively. The goal is to find a subset $s\subseteq v$, such that the number of edges between $s$ and its complement subset is maximized.\\

An objective (energy) function $f$ must use the MAC operation in converging to a solution. We use the framework of a \textit{Hopfield model} to achieve this (\onlinecite{hopfield1984neurons,cai2020powerHopfield}). In this scheme, we map the the graph problem to the energy function $
 f=-\frac{1}{2}\sum_{\text{mn}}^{\text{N}} -w_{\text{mn}}\nu_{\text{m}}\nu_{\text{n}}$ 
where $w_{\text{mn}}$ is the adjacency matrix that encodes the strength of the edges. $\nu_{\text{m}}$ and $\nu_{\text{n}}$ $\in \{-1, 1\}$ represent the state of interconnected nodes in a network of $N$ nodes. The energy has an arbitrary unit scale, which is a function of the graph problem (connection density and weight values), and the values of the binary nodal states denoted as $H$ and $L$. To find an optimal solution, the energy function is iteratively minimized by toggling the nodal states. This is done by using the update-rules: 
$\nu_m = \begin{cases}
        H & \text{if } \frac{\sum_{m\neq n}w_{\text{mn}}\nu_{\text{m}}}{\gamma} + \eta\geq \theta_{\text{ref}}\\
        L & \text{if } \frac{\sum_{m\neq n}w_{\text{mn}}\nu_{\text{m}}}{\gamma} + \eta<\theta_{\text{ref}}  
    \end{cases} $
for some threshold $\theta_{ref}$ and some noise value $\eta$ drawn from a normal distribution. The normalization constant $\gamma =\max_{v \in G}{\text{(D)}}$ equals the maximum degree $\text{D}$ of all graph vertices. This ensures the MAC result to stay in the interval [-1, 1]. When the optimization is performed, the system converges to the nodal state distributions that minimize the net energy. Each agent optimizes independently by picking random nodes and applying the update rules. In regular intervals the agents interact via crossover and each agent inherits a number of nodal states from the current best agent. The number of nodes inherited increases with time. The specific schedule is elaborated on in the methods section. The algorithm is outline in pseudocode as Algorithm \ref{alg:max_cut_alg}.\\

As an example, we solve a cyclic graph of 40 nodes. Briefly, the system is initialized with a random distribution of the nodal states, where some nodes belong to a set $Set_1$, and others to the complementary set $Set_2$. Each node can be in a state of $H$ or $L$. Furthermore, each node has edges with only immediate neighbors ($w_{\text{mn}}=w_{\text{nm}}$), and there are no self-loops ($w_{\text{mn}}=0$, if $m=n$).  Four random solution candidates representing individual agents (wavelengths) are initialized. We choose 0 as the threshold value $\theta_{ref}$.  Although the multi-agent approach is capable of optimal convergence, this typically necessitates an increase in both the number of agents and computational iterations. Combining them with photonic annealing, however, improves the convergence success rate. \\

We emulate different noise decay profiles and benchmark the performance against the experimentally obtained noise profiles and fully deterministic optimization (see Figure \ref{fig:5_max_cut}b-c). See methods for the rescaling method of the noise benchmark data. The injected noise profiles follow the form earlier described with $B=0$ and $A=\sqrt{\text{cycles}/\alpha}$, where we sweep the $\alpha$ parameter to obtain the different profiles, as shown in Figure \ref{fig:5_max_cut}c. These values dictate the standard deviation of the normal distribution used to draw the values $\eta$ in the update rules. Figure \ref{fig:5_max_cut}b shows the individual Hopfield energies of one example simulation using the experimental noise profiles. Each noise profile is executed 20 times and we record the average Hopfield global energy in Figure \ref{fig:5_max_cut}c. The optimal solution for this problem has an energy of -40. When noise is injected the solver reliably finds the correct solution regardless of noise magnitude within the emulated noise range. The chosen noise profile seem to affect convergence speed, with higher initial noise profiles delaying discovery of the optimal solution. However, profiles with low initial noise below $\approx 0.5$ show very similar behavior to each other. This suggests that the system is robust against high initial noise when agents are more in the exploratory phase, but indifferent to noise below some threshold. We also observe that the noise profiles extrapolated from data are among the best in terms of convergence speed. Thus, we expect the proposed IPO system to perform well on this problem based on these simulations.\\

To benchmark our approach we solve 30 non-circular graphs from the open sourced BiqMac-Library  (\onlinecite{biqmac}). We use 4 agents and the experimentally obtained noise profiles.  These problems cover undirected and unweighted graphs with edge probability of 50\% of sizes 60, 80 and 100 nodes with 10 examples each.  Each graph is solved multiple times, and the solutions are tabulated to compare the success rates. Already without any hyperparameter tuning, we are able to find the most optimal solutions of every graph (see bottom panel of Figure \ref{fig:5_max_cut}d), at least more than once. This is represented as success rates, which vary between 10\% and 100\% showing slightly worse rates for the graphs of large sizes. Despite the low success rate for some graphs, we note that many obtained solutions are within 1\% of the true solution, even when the algorithm does not fully converge. This is shown in the top panel of Figure \ref{fig:5_max_cut}d where we plot the distribution of the relative remaining error for all runs. We note that the success rates could be favorably improved by increasing the number of agents and further optimizing the noise profiles and nature of the inter-agent cooperation.  

\subsection*{Solving Continuous Optimization:}

\begin{figure}[h!]
    \includegraphics[width=0.9\textwidth]{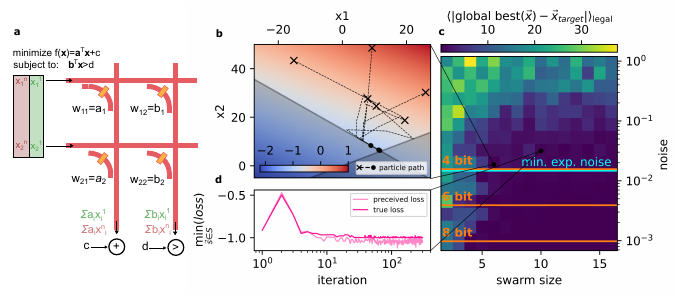}
    \caption{\textbf{Linear Programming Problems}. \textbf{(a)} A schematic showing the configuration of a crossbar unit used for solving a constrained linear optimization problem through canonical particle swarm approach. Multiple queries encoded into the amplitude modulated optical signals operate in parallel. Evaluation of constraints are uniquely handled in the same pass by mapping into extra columns in the crossbar. \textbf{(b)} A snapshot of 6 agents’ trajectories during balanced exploration and exploitation under moderate injected photonic noise in an emulated particle swarm optimization problem. The bottom panel shows evolution of the cost function as a function of iterations. \textbf{(c)} A plot showing the final average swarm position distance from the true minimum for canonical particle swarm problem for swarm sizes 2 to 16. The experimentally achievable noise levels and comparative approximate bit-accuracies are indicated.\textbf{(d)} The minimum swarm loss as a function of number of iterations for a swarm size 10 is shown. Here, the perceived loss is the loss recorded by the swarm through noisy evaluations, while true loss the exact loss at the resulting positions of each agent.} 
    \label{fig:4_PSO}
\end{figure}

\noindent  We now discuss the utility of the IPO in linear programming, i.e., solving optimization problems with linear objective functions and linear constraints. We discuss the solution to a constrained optimization problem. These are problems in which a function is to be minimized or maximized, subject to certain constraints. These problems are relevant in a range of applications, for example, in manufacturing where production is more generally constrained by resources and demands. In Figure \ref{fig:4_PSO}a we illustrate the mapping of a constrained optimization problem onto the compute core. Two equations describe the objective function and the constraint. To achieve parallel computations on both equations, we map the coefficients of the objective function onto the transmission states of the memory cells at the cross-points of one column and the constraint function to a second column.  We use PSO as the optimizer for this problem. In PSO, a collection of agents move throughout a region in discrete steps. After each iteration, the algorithm updates the velocity \(\vec{v}\) of all agents, nudging them towards the best solution (\onlinecite{shi2001particleswarm, wang2018particleswarm, zambrano-bigiarini_standard_2013}. We elaborate on the algorithmic specifics in the methods section and Algorithm~\ref{alg:pso_alg}.\\

In our proposed IPO both the loss and constraint evaluation are performed in-memory in the photonic crossbar array for all agents/particles in a single compute cycle. As an example, we evaluate the minimum of the constrained linear function, 
$f(x_1, x_2)=10x_1+20x_2,\text{ while }~3x_1+4x_2\geq60~\text{ and }~x_1-2x_2\leq0$. 
We normalize the loss landscape to the intervall [-1, 1], where -1 is the minimum within the constrained region and +1 the maximum based on the bound on the initial positions. We also constrain the area based on minimal and maximal values of the coordinates $x_1 \in [-27.5, 27.5]$ and $x_2 \in [-5, 50]$. Figure \ref{fig:4_PSO}b shows particle trajectories through the constrained landscape for a swarm size of 6 particles. Note, that all agents explore the solution landscape, constantly referencing other agents and adjusting their velocity based on their own experience and those of other agents. The constraints are implemented as absorptive walls meaning agents hitting a constraint are set to the boundary with their velocity set to zero  (\onlinecite{robinson_particle_2004, huang_hybrid_2005}). \\

We benchmark this implementation against increasing uncertainty in the evaluation of the loss function as well as the constraints. Given the continuous nature of the variables, the compute imprecision can be more deteriorating in linear programming. Here, noise affects optimization as it induces uncertainty in the quality of each agent's present and past positions. This is illustrated in Figure \ref{fig:4_PSO}d, where we plot the global minimum swarm loss over time of an example simulation using 10 particles. We see that the perceived minimum loss, i.e. the minimum loss the swarm recorded is lower than the true loss at the recorded positions. Thus, the estimates of the current best position are likely often slightly incorrect, pulling the swarm in a slightly wrong direction.\\

In addition, noisy constraint evaluation can lead to agentic positions being incorrectly accepted or rejected. To still obtain solutions within the bounds we apply a single noiseless boundary condition check on the personal best positions of all agents within the swarm, rejecting out of bounds agents at the end of each optimization. The distance to the optimum to the thus obtained best legal minimum is plotted in figure \ref{fig:4_PSO}c. We repeat each configuration for 10 different random initial particle positions and velocities and take the average over all runs that produced at least one minimum within bounds. We find that noise can be readily compensated by increasing the swarm size. Based on the minimal noise obtained from our benchmark experiment we expect our IPO to perform well on this problem with agent sizes larger than 5. As a reference we also mark noise values equivalent to $1/2 \cdot 1/2 \cdot 1/2^b$ where b is a certain bit resolution $b \in \{4, 6, 8\}$. The $2 \sigma$ interval of this standard deviation value equals half the level spacing of the respective bit-resolution. Thus, this value corresponds to a noise level equivalent to a $\approx 5\%$ bit-error probability on the least significant bit during discretization. Regardless, optimization problems needing high compute precision can benefit from stabilized light sources, such as mode-locked or narrowband lasers.

\section*{Discussion:}

\noindent Integrated photonic computing architectures can achieve significantly higher compute densities than digital electronics, owing to the speed of optics and the parallelism enabled by spatial multiplexing. When combined with non-volatile elements such as phase-change photonic memory devices, this approach can further facilitate fast, and efficient in-memory computing computational hardware (in supplementary information S1 we discuss IPO with PCM devices). Several optical architectures have also been demonstrated in the past few years for solving optimization problems. However, these are largely based on coherent networks that use degenerate optical parametric oscillators (\onlinecite{mcmahon2016fullyOPOoscillator}), injection-locked laser networks (\onlinecite{utsunomiya2015binaryLASER}) or spatial multiplexing (\onlinecite{pierangeli2019largeSpatialMuLEPEXING}). Working with time-multiplexed optical signals , these, therefore, do not capitalize on the parallelism property, and face scaling limitations due to dispersion and decoherence. More recently, integrated Mach–Zehnder interferometers were demonstrated for coherent networks. Such on-chip networks significantly reduce the footprint, however, they are volatile and require high standby power. The latter also set a limit on the size of such networks, since individual driver units must support individual interferometers \textcolor{black}{(in supplementary information S4, we compare these various different on-chip architectures)}. \\

In the ideal IPO, multiple MAC operations can be executed within sub-nanosecond latency, enabled by binary and pulse-amplitude input encoding (\textcolor{black}{see supplementary information S4}). The maximum achievable speed is ultimately limited by the speed of the modulators, photo-detectors and the bandwidths of the ADCs and DACs of the control board. Achieving comparable performance using conventional digital electronics remains extremely challenging. However, the estimation of the energy function and the corresponding feedback loop remain bottlenecks, as these steps are still executed electronically in each iteration. That said, this limitation is still less critical than in photonic neural network computing. Since our approach employs only a single layer and simple auxiliary operations, a high-bandwidth controller is sufficient to maintain the system in a compute-bound regime. \textcolor{black}{Furthermore, because the the optical hardware remains idle during electronic processing, we can enable static batching, where the same wavelengths are reused across time slots to represent different agents.  If $P$ wavelengths are available and $B$ agents are time-multiplexed, the total number of agents scales as $N_{\mathrm{agents}} = P \times B$. Since photonic latency is inherently low, increasing $B$ adds minimal overhead while significantly expanding parallel exploration}. Nonetheless, an ideal IPO would have simpler update rules, which could be realized entirely in the optical domain. Realizing IPO in fully integrated photonic hardware may also present certain challenges, one of which is input encoding. This limitation can be addressed by using broadband incoherent light sources, thereby removing the need for cascaded de-multiplexers. Moreover, this approach can leverage the inherent entropy source  (\onlinecite{bruckerhoff-pluckelmann_probabilistic_2024}), enabling random permutations to be incorporated into state updates. An IPO can also be envisioned with provisions for both deterministic and stochastic sources, or other noise injection mechanisms such as on the output nodes, allowing optimization problems that require high compute precision to also be solved with higher success rates. \textcolor{black}{We also point out that the numeric precision of compute depends on whether the optimization problem is discrete or continuous. For example, in discrete optimization problems small errors in the MVM can be tolerated as long as the sign (threshold crossing) is correct, making the operation robust to limited bit-width.}\\

In our proposed method of solving optimization problems, we take a population-based meta-heuristic approach. We implement in-memory search operations by leveraging the parallelism from WDM in conjunction with utilizing inherent hardware noise to our advantage.  \textcolor{black}{While the forward-pass operation relies on MAC operations, similar to neural network implementations on photonic circuits, we demonstrate that this hardware can function as an iterative solver for optimization tasks. Specifically, we treat the photonic substrate as a single-layer compute unit, enhanced with recurrent connections, and for discrete optimization problems, the data flow is reduced to a binary representation (\onlinecite{syed2025solving}). We also outline important algorithmic directions for implementing optimization problems. The use of multiple agents through dynamic referencing introduces an additional degree of freedom, enabling standard optimization algorithms to be realized more rigorously. Within this framework, information gain and entropy serve as principled metrics, which provide robust termination criteria}. Looking forward, unique schemes such as annealing can be distributed across agents and coordinated through feedback mechanisms, thereby shaping a collective annealing schedule. Similarly, \textcolor{black}{feedback-driven updates} can be employed to guide solutions according to evolutionary principles. Here the optimizer can track the evolutionary progress of individual agents as well as subsets of populations in real time, closely mirroring how these processes unfold in nature \textcolor{black}{(see Supplementary Section S1)}. Finally, while we demonstrate in preliminary simulations the end-to-end optimization of our proposed IPO unit it can also be of use in cases where it cannot solve the problem on it's own. The strong approximation capability for a good a solution can also enable the optimizer a valuable tool for initializing solutions (high-quality starting point compared to random initialization) in downstream annealing-based solvers (\onlinecite{Y1995goemansACM,Y2021EggerQuantum}).\\

\noindent In summary, we have presented a co-designed framework that combines photonic hardware with algorithmic strategies to tackle challenging optimization problems. We introduce an integrated photonic optimizer that leverages low-cost in-memory multiply-and-accumulate operations, implemented within a nature-inspired multi-agent cooperative scheme supported by multiplexing. As demonstrative examples, we use experimentally benchmarked hardware to emulate solutions for both continuous and discrete graph problems. By performing computation where the analog data resides and enabling truly parallel multi-agent processing, this approach overcomes the energy and latency limitations inherent to electronic systems. Furthermore, the approach exploits intrinsic hardware noise, transforming it into a computational resource of optimization problems.


\section*{References}

\def\url#1{}
\bibliographystyle{ieeetr}

\begin{thebibliography}{10}

\bibitem{kennedy1995particle}
J.~Kennedy and R.~Eberhart, ``Particle swarm optimization,'' in {\em Proceedings of ICNN'95-international conference on neural networks}, vol.~4, pp.~1942--1948, IEEE, 1995.

\bibitem{dorigo2006ant}
M.~Dorigo, M.~Birattari, and T.~Stutzle, ``Ant colony optimization,'' {\em IEEE computational intelligence magazine}, vol.~1, no.~4, pp.~28--39, 2006.

\bibitem{coopertaiveschloesser2021individual}
D.~S. Schloesser, D.~Hollenbeck, and C.~T. Kello, ``Individual and collective foraging in autonomous search agents with human intervention,'' {\em Scientific Reports}, vol.~11, no.~1, p.~8492, 2021.

\bibitem{torney2011signallingcooperative}
C.~J. Torney, A.~Berdahl, and I.~D. Couzin, ``Signalling and the evolution of cooperative foraging in dynamic environments,'' {\em PLoS computational biology}, vol.~7, no.~9, p.~e1002194, 2011.

\bibitem{dafoe2021cooperative}
A.~Dafoe, Y.~Bachrach, G.~Hadfield, E.~Horvitz, K.~Larson, and T.~Graepel, ``Cooperative ai: machines must learn to find common ground,'' {\em Nature}, vol.~593, no.~7857, pp.~33--36, 2021.

\bibitem{gupta2017cooperative}
J.~K. Gupta, M.~Egorov, and M.~Kochenderfer, ``Cooperative multi-agent control using deep reinforcement learning,'' in {\em Autonomous Agents and Multiagent Systems: AAMAS 2017 Workshops, Best Papers, S{\~a}o Paulo, Brazil, May 8-12, 2017, Revised Selected Papers 16}, pp.~66--83, Springer, 2017.

\bibitem{reinforcetan1993multi}
M.~Tan, ``Multi-agent reinforcement learning: Independent vs. cooperative agents,'' in {\em Proceedings of the tenth international conference on machine learning}, pp.~330--337, 1993.

\bibitem{booktalbi2009metaheuristics}
E.-G. Talbi, {\em Metaheuristics: from design to implementation}.
\newblock John Wiley \& Sons, 2009.

\bibitem{surveygogna2013metaheuristics}
A.~Gogna and A.~Tayal, ``Metaheuristics: review and application,'' {\em Journal of Experimental \& Theoretical Artificial Intelligence}, vol.~25, no.~4, pp.~503--526, 2013.

\bibitem{blum2003metaheuristics}
C.~Blum and A.~Roli, ``Metaheuristics in combinatorial optimization: Overview and conceptual comparison,'' {\em ACM computing surveys (CSUR)}, vol.~35, no.~3, pp.~268--308, 2003.

\bibitem{sebastian2020memoryAbu}
A.~Sebastian, M.~Le~Gallo, R.~Khaddam-Aljameh, and E.~Eleftheriou, ``Memory devices and applications for in-memory computing,'' {\em Nature nanotechnology}, vol.~15, no.~7, pp.~529--544, 2020.

\bibitem{zhou2009GPU}
Y.~Zhou and Y.~Tan, ``Gpu-based parallel particle swarm optimization,'' in {\em 2009 IEEE Congress on Evolutionary Computation}, pp.~1493--1500, IEEE, 2009.

\bibitem{chakroun2013combiningGPU}
I.~Chakroun, N.~Melab, M.~Mezmaz, and D.~Tuyttens, ``Combining multi-core and gpu computing for solving combinatorial optimization problems,'' {\em Journal of Parallel and Distributed Computing}, vol.~73, no.~12, pp.~1563--1577, 2013.

\bibitem{tsukamoto2017acceleratorFPGA}
S.~Tsukamoto, M.~Takatsu, S.~Matsubara, and H.~Tamura, ``An accelerator architecture for combinatorial optimization problems,'' {\em Fujitsu Sci. Tech. J}, vol.~53, no.~5, pp.~8--13, 2017.

\bibitem{Y2021gpuzhang}
F.~Zhang, C.~Zhao, S.~Han, F.~Ma, and D.~Xiang, ``Gpu-based parallel implementation of vlbi correlator for deep space exploration system,'' {\em Remote Sensing}, vol.~13, no.~6, p.~1226, 2021.

\bibitem{mutlu2022modern}
O.~Mutlu, S.~Ghose, J.~G{\'o}mez-Luna, and R.~Ausavarungnirun, ``A modern primer on processing in memory,'' in {\em Emerging Computing: From Devices to Systems: Looking Beyond Moore and Von Neumann}, pp.~171--243, Springer, 2022.

\bibitem{Y2019computationalsebastian}
A.~Sebastian, M.~Le~Gallo, and E.~Eleftheriou, ``Computational phase-change memory: Beyond von neumann computing,'' {\em Journal of Physics D: Applied Physics}, vol.~52, no.~44, p.~443002, 2019.

\bibitem{rios2015integrated}
C.~R{\'\i}os, M.~Stegmaier, P.~Hosseini, D.~Wang, T.~Scherer, C.~D. Wright, H.~Bhaskaran, and W.~H. Pernice, ``Integrated all-photonic non-volatile multi-level memory,'' {\em Nature photonics}, vol.~9, no.~11, pp.~725--732, 2015.

\bibitem{wuttig2017phase}
M.~Wuttig, H.~Bhaskaran, and T.~Taubner, ``Phase-change materials for non-volatile photonic applications,'' {\em Nature photonics}, vol.~11, no.~8, pp.~465--476, 2017.

\bibitem{Y2021feldmannparallelconvo}
J.~Feldmann, N.~Youngblood, M.~Karpov, H.~Gehring, X.~Li, M.~Stappers, M.~Le~Gallo, X.~Fu, A.~Lukashchuk, A.~S. Raja, {\em et~al.}, ``Parallel convolutional processing using an integrated photonic tensor core,'' {\em Nature}, vol.~589, no.~7840, pp.~52--58, 2021.

\bibitem{ghazi2022integrated}
S.~Ghazi~Sarwat, F.~Br{\"u}ckerhoff-Pl{\"u}ckelmann, S.~G.-C. Carrillo, E.~Gemo, J.~Feldmann, H.~Bhaskaran, C.~D. Wright, W.~H. Pernice, and A.~Sebastian, ``An integrated photonics engine for unsupervised correlation detection,'' {\em Science Advances}, vol.~8, no.~22, p.~eabn3243, 2022.

\bibitem{hua2025integrated}
S.~Hua, E.~Divita, S.~Yu, B.~Peng, C.~Roques-Carmes, Z.~Su, Z.~Chen, Y.~Bai, J.~Zou, Y.~Zhu, {\em et~al.}, ``An integrated large-scale photonic accelerator with ultralow latency,'' {\em Nature}, vol.~640, no.~8058, pp.~361--367, 2025.

\bibitem{dong2024partial}
B.~Dong, F.~Br{\"u}ckerhoff-Pl{\"u}ckelmann, L.~Meyer, J.~Dijkstra, I.~Bente, D.~Wendland, A.~Varri, S.~Aggarwal, N.~Farmakidis, M.~Wang, {\em et~al.}, ``Partial coherence enhances parallelized photonic computing,'' {\em Nature}, vol.~632, no.~8023, pp.~55--62, 2024.

\bibitem{Y2019fastXuanli}
X.~Li, N.~Youngblood, C.~R{\'\i}os, Z.~Cheng, C.~D. Wright, W.~H. Pernice, and H.~Bhaskaran, ``Fast and reliable storage using a 5 bit, nonvolatile photonic memory cell,'' {\em Optica}, vol.~6, no.~1, pp.~1--6, 2019.

\bibitem{fortier_20_2019}
T.~Fortier and E.~Baumann, ``20 years of developments in optical frequency comb technology and applications,'' {\em Communications Physics}, vol.~2, p.~153, Dec. 2019.
\newblock Publisher: Nature Publishing Group.

\bibitem{bruckerhoff-pluckelmann_probabilistic_2024}
F.~Brückerhoff-Plückelmann, H.~Borras, B.~Klein, A.~Varri, M.~Becker, J.~Dijkstra, M.~Brückerhoff, C.~D. Wright, M.~Salinga, H.~Bhaskaran, B.~Risse, H.~Fröning, and W.~Pernice, ``Probabilistic photonic computing with chaotic light,'' {\em Nature Communications}, vol.~15, p.~10445, Dec. 2024.
\newblock Publisher: Nature Publishing Group.

\bibitem{Y1948shannonmathematical}
C.~E. Shannon, ``A mathematical theory of communication,'' {\em The Bell system technical journal}, vol.~27, no.~3, pp.~379--423, 1948.

\bibitem{van1987simulatedannealing}
P.~J. Van~Laarhoven, E.~H. Aarts, P.~J. van Laarhoven, and E.~H. Aarts, {\em Simulated annealing}.
\newblock Springer, 1987.

\bibitem{muhlenbein1988evolution}
H.~M{\"u}hlenbein, M.~Gorges-Schleuter, and O.~Kr{\"a}mer, ``Evolution algorithms in combinatorial optimization,'' {\em Parallel computing}, vol.~7, no.~1, pp.~65--85, 1988.

\bibitem{katoch2021reviewevolution}
S.~Katoch, S.~S. Chauhan, and V.~Kumar, ``A review on genetic algorithm: past, present, and future,'' {\em Multimedia Tools and Applications}, vol.~80, pp.~8091--8126, 2021.

\bibitem{shi2001particleswarm}
Y.~Shi {\em et~al.}, ``Particle swarm optimization: developments, applications and resources,'' in {\em Proceedings of the 2001 congress on evolutionary computation (IEEE Cat. No. 01TH8546)}, vol.~1, pp.~81--86, IEEE, 2001.

\bibitem{wang2018particleswarm}
D.~Wang, D.~Tan, and L.~Liu, ``Particle swarm optimization algorithm: an overview,'' {\em Soft computing}, vol.~22, pp.~387--408, 2018.

\bibitem{Y2003MAXCUT}
J.-Y. Potvin and K.~A. Smith, ``Artificial neural networks for combinatorial optimization,'' in {\em Handbook of metaheuristics}, pp.~429--455, Springer, 2003.

\bibitem{Y2011combinatorialMAXcut}
B.~H. Korte, J.~Vygen, B.~Korte, and J.~Vygen, {\em Combinatorial optimization}, vol.~1.
\newblock Springer, 2011.

\bibitem{hopfield1984neurons}
J.~J. Hopfield, ``Neurons with graded response have collective computational properties like those of two-state neurons.,'' {\em Proceedings of the national academy of sciences}, vol.~81, no.~10, pp.~3088--3092, 1984.

\bibitem{cai2020powerHopfield}
F.~Cai, S.~Kumar, T.~Van~Vaerenbergh, X.~Sheng, R.~Liu, C.~Li, Z.~Liu, M.~Foltin, S.~Yu, Q.~Xia, {\em et~al.}, ``Power-efficient combinatorial optimization using intrinsic noise in memristor hopfield neural networks,'' {\em Nature Electronics}, vol.~3, no.~7, pp.~409--418, 2020.

\bibitem{biqmac}
A.~Wiegele, ``Biq {Mac} {Library} - {BInary} {Quadratic} and {MAx} {Cut} {Library},'' 2007.

\bibitem{zambrano-bigiarini_standard_2013}
M.~Zambrano-Bigiarini, M.~Clerc, and R.~Rojas, ``Standard {Particle} {Swarm} {Optimisation} 2011 at {CEC}-2013: {A} baseline for future {PSO} improvements,'' in {\em 2013 {IEEE} {Congress} on {Evolutionary} {Computation}}, pp.~2337--2344, June 2013.
\newblock ISSN: 1941-0026.

\bibitem{robinson_particle_2004}
J.~Robinson and Y.~Rahmat-Samii, ``Particle swarm optimization in electromagnetics,'' {\em IEEE Transactions on Antennas and Propagation}, vol.~52, pp.~397--407, Feb. 2004.

\bibitem{huang_hybrid_2005}
T.~Huang and A.~Mohan, ``A hybrid boundary condition for robust particle swarm optimization,'' {\em IEEE Antennas and Wireless Propagation Letters}, vol.~4, pp.~112--117, 2005.

\bibitem{mcmahon2016fullyOPOoscillator}
P.~L. McMahon, A.~Marandi, Y.~Haribara, R.~Hamerly, C.~Langrock, S.~Tamate, T.~Inagaki, H.~Takesue, S.~Utsunomiya, K.~Aihara, {\em et~al.}, ``A fully programmable 100-spin coherent ising machine with all-to-all connections,'' {\em Science}, vol.~354, no.~6312, pp.~614--617, 2016.

\bibitem{utsunomiya2015binaryLASER}
S.~Utsunomiya, N.~Namekata, K.~Takata, D.~Akamatsu, S.~Inoue, and Y.~Yamamoto, ``Binary phase oscillation of two mutually coupled semiconductor lasers,'' {\em Optics express}, vol.~23, no.~5, pp.~6029--6040, 2015.

\bibitem{pierangeli2019largeSpatialMuLEPEXING}
D.~Pierangeli, G.~Marcucci, and C.~Conti, ``Large-scale photonic ising machine by spatial light modulation,'' {\em Physical review letters}, vol.~122, no.~21, p.~213902, 2019.

\bibitem{syed2025solving}
G.~S. Syed and A.~Sebastian, ``Solving optimization problems with photonic crossbars,'' Sept.~23 2025.
\newblock $\uppercase{US}$ Patent 12,422,882.

\bibitem{Y1995goemansACM}
M.~X. Goemans and D.~P. Williamson, ``Improved approximation algorithms for maximum cut and satisfiability problems using semidefinite programming,'' {\em Journal of the ACM (JACM)}, vol.~42, no.~6, pp.~1115--1145, 1995.

\bibitem{Y2021EggerQuantum}
D.~J. Egger, J.~Mare{\v{c}}ek, and S.~Woerner, ``Warm-starting quantum optimization,'' {\em Quantum}, vol.~5, p.~479, 2021.

\end{thebibliography}

\section*{Supplementary Materials}
\noindent Supplementary material for this article is available at....\\
\\
Section S1. Phase Change Memory based Weights.\\
Section S2. Further discussion on Linear Programming Implementation.\\
Section S3. Precision of Compute.\\
Section S4. Architecture Analysis.\\
Fig. S1. Combinatorial Optimization Problems.\\
Fig. S2. Linear Programming Problems.\\
Fig. S3. Success rate and stability of the PSO solutions.\\
Fig. S4. Iterative Solver.\\
Fig. S5. Photonic Crossbar Efficiency.\\
Fig. S6. Latency and Throughput Projections.\\
Table. S1. Performance comparison of representative computing platforms.\\

\clearpage

\section*{Acknowledgments}

\noindent S.G.S and W.P acknowledge NEQIOS project funded by the European Union’s Horizon Europe (grant agreement no. 101259183). S.G.S acknowledges partial funding from the European Research Council (grant agreement no. 101222715).  A. Sebastian acknowledges partial funding from the European Research Council (grant agreement no. 101201020).  W.P. acknowledges partial funding from the European Research Council (grant agreement no. 101200429). W.P. also acknowledges partial funding from DARPA NAPSAC program.

\section*{Competing interests}

\noindent The authors declare no competing financial interests.

\section*{Data, Code, and Materials}
\noindent This study did not generate new materials. The data that support the findings of this study are uploaded to a public repository (10.5281/zenodo.19660492).

\section*{Contributions}
\noindent G.S.S. conceived the research question and designed the algorithms and simulations. P.S. performed the hardware experiments and conducted the emulation studies. F.B.P. designed the experimental setup, which was built by J.D, and performed the initial hardware experiments. W.P. and A.S.E. provided technical guidance and management support. G.S.S wrote the manuscript with inputs from P.S. All authors read the manuscript.

\section*{Methods}

\small

\noindent\paragraph{Scaling experimental Noise}:
Our performance benchmark experiment consists of 1280 random MVMs of the type $\vec{y} = M\vec{x}, ~\vec{x} \in [-1, 1]\,{\mathbb{R}^{64}}, M\in[1]\,{\mathbb{N}^{64\times1}}$. Each vector component is drawn from a uniform distribution. This operation is then averaged to reduce hardware noise stemming from the ASE source and the readout electronics, in order to obtain the noise profiles shown in Figure~\ref{fig:3_msm_setup}. Supplementary information Figure S5a shows the resulting scatter plot of the target MVM operation versus the measured MVM result for two different averaging factors. As indicated in the figure, the MVM values lie approximately within the interval $[-15, 15]$. To obtain realistic noise magnitude estimates for different algorithms—where the results of the operations can vary significantly—we rescale these results. From the distribution of target operations, we choose the interval $[-3\sigma, 3\sigma]$ to correspond to $[-1, 1]$. From these normalized values, we obtain error distributions of Gaussian shape. These distributions narrow with increased averaging and are shown as heatmaps for all four measured channels in Figure S5b. We fit these distributions to extract the standard deviations for each averaging factor and then fit a $A/\sqrt{n} + B$ profile to the resulting values (see Figure S5c). The data reveal that the error distributions exhibit a slight deviation from zero mean, suggesting the presence of residual systematic error. We do not account for this offset when simulating our noise levels, assuming it can be mitigated through improved calibration in future experiments. As mentioned in the main text, all values with added noise are normalized to the same range of $[-1, 1]$, allowing us to directly transfer our normalized noise values into the simulations.\\

\noindent\paragraph{max‑cut solver}
The Max-Cut solver is initialized by assigning each agent a random cut of half the graph size. Each agent then optimizes independently by picking a random node and updates for 10 iterations. This concludes one optimization cycle after which the agents share solutions by inheriting a number of node configurations from the currently best performing agent, based on its Hopfield energy. The number of inherited nodes increases with cycles. The case of the circular graph we use 2000 cycles and share no nodes for the first 50 cycles, 25\% of the nodes from cycle 50 to 100 and 50\% of the nodes for all remaining cycles. Note that the shared nodes are chosen randomly, hence we are not necessarily sharing connected subgraphs. For the non-circular graphs, we also use 2000 cycles of length 10 and increase the cooperation every 400 cycles in 20\% steps starting with 0\% of nodes shared and ending at 80\% nodes shared. For the emulated noise profiles all agents share the same decay traces, and the four measured channels are distributed to the four agents. 

\begin{algorithm}[H]
\caption{Photonic Max‑Cut Solver}\label{alg:max_cut_alg}
\begin{algorithmic}[1]
\Require Graph \(G(V, E)\), objective function \(f_{{Hop}}: \textbf{s}^n \to \mathbb{R}\), number of agents \(n\), noise profile \(\eta(t) = \mathcal{N}(\mu,\,\sigma_t^2)\), number of cycles $n$, cycle length $m$
\State Initialize each agent’s solution \(\vec{s}^i \in \textbf{s}^n\) and compute initial energy \(E_i \leftarrow f_{\text{Hop}}(\vec{s}^i)\)
\For{n cycles}
    \For{k iterations}
    \State \textbf{(Energy Computation)} All agents compute simultaneously: agent \textit{i}
        \State pick random node $g_i$
        \State \(g'_i \leftarrow \text{MAC}(\vec{s}^i) + \eta(t)\) \Comment{Update node $g_i$} 
        
        \State \(s^{i'} \leftarrow \text{UpdateState}(\vec{s}^i, g'_i) \) \Comment{Update Graph State} 
        \State \(E_i' \leftarrow f_{\text{Hop}}(\vec{s}^i,g_i')\)  
    \EndFor
  \State \textbf{(Communication)}
        \State \(\vec{s}^c \leftarrow \arg\min_{\vec{s} \in S^n} f_{\text{Hop}}(\vec{s})\) 
        \State  Increase \(\leftarrow  \text{Similarity}(\vec{s}^i, \vec{s}^c) \)
\EndFor
    \State Determine the best state among all agents
\State \Return best solution 
\end{algorithmic}
\end{algorithm}

\noindent\paragraph{cont. optimization}
We use canonical PSO with absorptive bounds as described in references  (\onlinecite{robinson_particle_2004, huang_hybrid_2005}) to implement particle swarm optimization.  We use an adaptive random topology of size 3, meaning each particle informs at most 3 other particles beyond itself and shuffles these whenever a cycle did not improved the global optimum  (\onlinecite{zambrano-bigiarini_standard_2013}). 
The algorithm is outlined in references  (\onlinecite{zambrano-bigiarini_standard_2013, wang2018particleswarm}), and we briefly discuss it here. 
The velocity update equation for an agent $\vec{s}_i$ at iteration $t+1$,  follows $
    \vec{v}_i^{t+1} = \omega\vec{v}_i^t + c_1 \vec{u}_1^t\odot(\vec{p}_i^t - \vec{s_i}^t)  + c_2 \vec{u}_2^t\odot(\vec{l}_i^t - \vec{s}_i^t)$, where $\vec{p}^t$ is the current personal best position encountered by the particle and $\vec{l}^t$ the current local best within it's neighborhood. $\vec{u}_1^t$ and $\vec{u}_2^t$ are independent random vector drawn from a uniform distribution between [0, 1]. Note that $\odot$ is used to mark element wise multiplication. 
The hyperparameter constants are set to $\omega = 1/(2ln(2))$ and $c_1 = c_2 = 0.5 + ln(2)$. These are directly taken from reference  (\onlinecite{zambrano-bigiarini_standard_2013}).  As mentioned before, bounds are implemented as absorptive walls, i.e. a particle about to break a constraint is set to the intersection of it's velocity vector and the boundary with velocity 0  (\onlinecite{robinson_particle_2004, huang_hybrid_2005}).
For a legal new position the particle is updated by simply adding the new velocity to the old position:
$\vec{s}_i^{t+1} = \vec{s}_i^t + \vec{v}_i^{t+1}$.\\

We normalize the loss function to be -1 at the minimum and +1 at the maximum within the allowed initialization and computational bounds.
For our example case we set the initialization intervals to $x_1 \in [-27.5, 27.5]$ and $x_2 \in [-5, 50]$. We sample random points in these intervals, and resample for any coordinates that break any constraint until all particles are initialized at a legal position. We let each swarm develop for 300 iterations.
The bound values are normalized to the interval $[-1, 1]$ within this initialization area. 
To emulate noisy MAC-operations we draw a sample from a standard normal distribution scaled by the specific noise value of the run for both constraint and loss function evaluation.
Due to noisy constraint evaluation it is possible that we obtain positions outside of the constraints. 
Thus, to still obtain a valid final result we we sample the best recorded position of each particle and perform one noiseless constraint evaluation, rejecting out of bounds positions and returning the best legal position as the final result. This can lead to rejection of of all minima turning the run unsuccessful. Supplementary Section S2 shows the number of successful runs across the noise vs. swarm size sweep shown in Figure \ref{fig:4_PSO}.\\

\begin{algorithm}[H]

\caption{Parallel Wavelength Exploration (PSO)}\label{alg:pso_alg}
\begin{algorithmic}[2]
\Require Objective function \( f: \textbf{s} \to \mathbb{R}^m \), constraints \( r \), number of agents \( N \), neighborhood size \( K \), number of iterations \( n \)
\State Randomly initialize each agent’s solution \( \vec{s}^i \in S \) and compute \( f_i = f(\vec{s}^i) \), randomly select \( K \) other agents to inform (duplicates allowed).
\For{\(k\) iterations}
    \State \textbf{(Parallel compute)} All agents operate simultaneously:
        \State get personal best \(\vec{p}^i\)
        \State get local best \(\vec{l}^i\) from informants
        \State \(\vec{s}^{i'} \gets \text{Update}(\vec{s}^i, \vec{p}^i, \vec{l}^i)\)
        \If{\(s^{i'}\) satisfies \(r\)}
            \State \(\vec{s}^i, f_i \gets \vec{s}^{i'}, f(\vec{s}^{i'})\)
        \EndIf
        \State compute new global optimum \(\vec{g}'\)
        \If{\(\vec{g}'=\vec{g}\)}
            \State reshuffle neighborhoods
        \EndIf
\EndFor
\State \Return \( \vec{s}^j ~|~ f(\vec{s}^j) = \min_{\vec{s}^i \in Swarm} f(\vec{s}^i) \)
\end{algorithmic}
\end{algorithm}

\noindent\paragraph{Experimental setup}:
The ASE lightsource (Agilent 83438A) is spectrally filtered by a wavelength division multiplexer and subsequently amplified by two amplifier stages (Pritel LNHPFA-30 pre-Amp into Pritel FA-33-IO-1807-23-009 power Amp) to approx. \unit[100]{mW} average power. The core of the IPO consists of a photonic integrated circuit containing a 9$\times$3 photonic crossbar array with built-in modulators and photodiodes. The input weights are encoded by 9 electric absorption modulators same as the 27
crossbar weights. Only 5 input lanes and 2 output lanes (i.e., a 5x2 subsection) of the crossbar are used in the experiment. The output of the on-chip photodiodes is amplified by off-chip transimpedance amplifiers (FEMTO HSA-Y-1-60) and fed to an FPGA. Said FPGA (Xilinx HW-Z1-ZCU216-REVA03) controls the IPO by setting
matrix and input weights as well as handling the data streams to an from the circuit via a custom Python API. The DACs driving the input EAMs operate at 4 \unit{GSa/s}, the ADCs reading the photodiode currents operate at 2 \unit{GSa/s}. Every input symbol gets expanded to four DAC samples and two consecutive ADC samples are combined into one symbol, resulting in a potential symbol rate of 1 \unit{GBd}. To encode negative numbers balanced detection of the 2
crossbar outputs is performed in the FPGA reducing the matrix dimension to 1$\times$5. Each
modulator is calibrated to take crosstalk effects and non-linear weight scaling
into account, calibration has been performed using the C34 input channel. The Output signal is scaled for every matrix according to an
empirically determined rescaling factor based on an example MVM with 1280 random
input vectors at an averaging factor of $2^8$. The noise profiles used for simulating experimental noise levels were obtained from performing MVMs of 1280 random input vectors size 64 with a constant matrix of all weights equal to one. \textcolor{black}{In supplementary information section S4, the power consumption in different sections of the demonstrator system is discussed.}

\onecolumngrid

\normalsize

\section*{Figures}

\clearpage
\newpage

\end{document}